\documentclass[longbibliography,aps,prx,twocolumn]{revtex4-1}

\usepackage[utf8]{inputenc}
\usepackage[T1]{fontenc}
\usepackage{lineno}
\usepackage{soul}
\usepackage{graphicx}
\usepackage{amsmath}
\usepackage{wrapfig}
\usepackage{csquotes}
\usepackage[english]{babel}

\usepackage{setspace}

\usepackage{comment}

\usepackage[utf8]{inputenc}
\usepackage{newunicodechar}

\newcommand{\fig}{Fig.~\ref}

\begin{document}

\title{Collective behavior of ``flexicles''}

\author{Philipp W. A. Sch\"onh\"ofer$^1$ and Sharon C. Glotzer$^{1,2}$}%
\email{Philipp W. A. Sch\"onh\"ofer (pschoenh@umich.edu), Sharon C. Glotzer (sglotzer@umich.edu)}
\affiliation{Department of Chemical Engineering, University of Michigan, Ann Arbor, Michigan 48109, USA.}
\affiliation{Biointerfaces Institute, University of Michigan, Ann Arbor, Michigan 48109, USA.}

%\affil[$\dag$]{these authors contributed equally to this work}

\begin{abstract}
\normalsize
In recent years the functionality of synthetic active microparticles has edged even closer to that of their biological counterparts. However, we still lack the understanding needed to recreate at the microscale key features of autonomous behavior exhibited by microorganisms or swarms of macroscopic robots. In this study, we propose a model for a three-dimensional deformable cellular composite particle consisting of self-propelled rod-shaped colloids confined within a flexible vesicle - a superstructure we call a ``flexicle''. Using molecular dynamics simulations, we investigate the collective behavior of dense systems comprised of many flexicles. We show that individual flexicles exhibit shape changes upon collisions with other flexicles that lead to rearrangement of the internal active rods that slow the flexicle motion significantly. This shape deformability gives rise to a diverse set of motility-induced phase separation phenomena and the spontaneous flow of flexicles akin to the migration of cells in dense tissues. Our findings establish a foundation for designing responsive cell-like active particles and developing strategies for controlling swarm migration and other autonomous swarm behaviors at cellular and colloidal scales. 
\end{abstract}
% \begin{document}

\maketitle
\section{Introduction}

In the field of active matter \cite{ND2017,R2017}, where individual building blocks of a system can harness external energy to perform work, there is a growing aspiration to find strategies for efficiently and precisely controlling active particles and collective motion to perform complex tasks. Various ideas for navigating swarms of self-propelled particles have been realized experimentally or proposed computationally. Although these strategies show great potential in the fields of medicine and engineering \cite{ZWLDS2019,LYTGWS2023}, most rely on direct guidance by magnetic, optical, electric or acoustic fields \cite{LYTGWS2023,WCXLZY2024} or require intricate particle interactions which limit their functionality \cite{O'KHS2017,MCDERLCMMcEC2020,PM2020,YB2020}. At the same time, common synthetic self-propelling particles like Janus particles \cite{WM2013} and Quincke rollers \cite{BCDDB2013} are rigid bodies and can only reproduce some behaviors observed in nature, such as motility-induced phase separation (MIPS), where the self-propelled agents spontaneously form dense clusters within a low-density gas phase due to their density-dependent velocities \cite{CT2015}. Biological cells, however, are flexible and can adapt to their local environments morphologically and dynamically. It has been shown that reconfiguring the active agent’s shape, controlling the motion of active elements inside the cell, and controlling cell rigidity are important for more complex phenomena \cite{BBC2024} such as pseudopodia \cite{PMGDSBCCPPRVMWDBA2018,vH2011}, the elongation of amoeba as a response to compression stresses \cite{RNSLL1999, CCTCN2013}, cell migration \cite{BYMM2016,MM2018}, embryonic development \cite{SLTM2013,KPS-VC2021} and cancer growth \cite{PFGATRKOZMK2015}.

Drawing inspiration from these flexible, autonomous biological objects, researchers have begun exploring active particles restricted to surfaces \cite{WL2008,YLM2014,NG-SB2019,VMM2023} or enclosed within confined spaces \cite{C-VS2018,SG2022,IWFG2023}. While most of these studies involve rigid confinement, recently conducted experiments and simulations have shown that self-propelled agents inside flexible vesicular membranes \cite{VHA-VvBDAFGV2020,PBH2021,IGF2022,PLG2022,SFUFPVBHB2023} or droplets \cite{XGCARK2022} can either induce or inhibit morphological changes to their deformable confinement. Active filaments can also act as actuators, clustering at the membrane interface, exerting localized pressure on the membrane wall, and generating second-order self-propelled vesicle particles. Although initially introduced in macroscopic robotic swarm systems \cite{BLLBDXCTBBK2021}, this mechanism has only recently been theoretically predicted for two-dimensional microscopic systems \cite{LSG2023,UHB2023} and can be seen as an alternative to vesicles and droplets propelled via liquid-liquid phase separation \cite{WBFWDBA2023,J-PTLSRD2024}. Additionally, initial experiments on the colloidal scale involving Janus particles engulfed in giant vesicles \cite{LNBDMPS2022} and Quincke rollers suspended in a liquid droplet \cite{KFPDV2022} have been reported. However, these studies focus solely on the individual motion of the confining object and do not address the collective behavior of systems of agents, which is crucial for identifying self-organization strategies. It is understood that the shape of particles, in the form of active colloidal polygons or polyhedra, influences MIPS behavior and the internal dynamics of clusters both in single-component systems \cite{MBSG2022} and binary mixtures \cite{MSG2022}. Furthermore, phase field calculations involving shape-deformable particles have revealed the suppression of MIPS in two dimensions \cite{HLCMM2023}. The active flexicle system investigated here introduces another layer of complexity, as the flexicles can not only change their shape in response to particle interactions and dense arrangements but also influence the internal dynamics of neighboring flexicles, producing novel, collective behaviors at the system level.

In this paper, we investigate via molecular dynamics simulations how shape flexibility and membrane rigidity influence the collective dynamics and clustering behavior of active-particle-driven vesicles or flexicles. To achieve this, we introduce a minimal model for a three-dimensional morphologically flexible cellular object composed of rigid self-propelled rods confined inside a vesicle; a composite particle we term \textit{flexicle} (see \fig{fig:method}). Similar to their two-dimensional counterparts \cite{LSG2023,UHB2023}, flexicles can undergo locomotion and act as individual deformable active agents. We construct a ``phase'' diagram of steady-state behaviors for a dense multi-flexicle system. In addition to a disordered fluid phase, we report that flexicles either enter a MIPS state involving both local density and particle shape (combined MIPS) or a MIPS state equivalent to that observed for hard particles \cite{CT2015} without significant morphological changes to the vesicles (singular MIPS), depending on membrane rigidity. Furthermore, we observe a jamming transition, above which flexicles with low deformability move collectively in persistent stacks or ``rouleaux'' that promote flexicle alignment.

\section{Methods}
\begin{figure}[t!]
\centering
\includegraphics[width=0.42\textwidth]{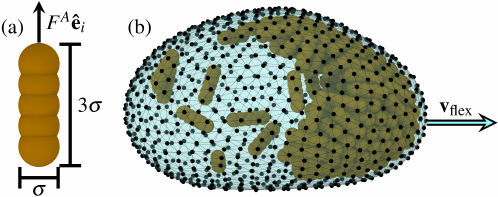}
\label{fig:method}
\caption{Computational model of a self-propelled rod particle (a) and vesicular flexicle (b). Each rod is comprised of 5 rigidly linked, overlapping spheres with diameter $\sigma$. The membrane of each flexicle is modeled by a set of vertices connected via triangulated mesh. The flexicle is propelled by the rods, which spontaneously form a polar cluster and push it in direction $\hat{\mathbf{v}}_\text{flex}$. }
\end{figure}

We performed molecular dynamics simulations of $N_\text{rod}$ self-propelled rigid rod-shaped particles of width $\sigma$ and aspect ratio $\alpha=4$ confined inside 3-dimensional vesicles of radius $R=8\sigma$. The system is contained within a cubic box with periodic boundary conditions and side length $L$. The vesicle is modeled as an oriented, triangulated mesh $\mathcal{M}$ with $N_v=900$ vertices and $N_t=1796$ triangles (see \fig{fig:method}). Vertices $i$ and $j$ at position $\mathbf{r}_i$ and $\mathbf{r}_j$ that share a common edge $e\in\mathcal{M}$ of length $l_{ij}=|\mathbf{r}_i-\mathbf{r}_j|$ are linked via a tether potential
\begin{equation}
U_\text{T}(l_{ij}) =
\begin{cases}
\kappa_\text{T}\frac{\text{exp}(1/(l_{ij}-l_0))}{l_{ij}-l_\text{min}} &\text{if $l_{ij}<l_0$}\\
\kappa_\text{T}\frac{\text{exp}(1/(l_1-l_{ij}))}{l_\text{max}-l_{ij}} &\text{if $l_{ij}>l_1$}\\
0 & \text{else.}
\end{cases}
\end{equation}
where $\kappa_\text{T}$ is the tether strength. This potential guarantees that $l_{ij}$ lies within minimum and maximum bond lengths $l_\text{min}=\frac{2}{3}\sigma$ and $l_\text{max}=\frac{4}{3}\sigma$ and that the mesh vertices move freely between the cutoff lengths $l_1=0.85\sigma$ and $l_0=1.15\sigma$. To model the bending rigidity we impose a bending potential between all neighboring triangles $t_a$ and $t_b$ with a shared edge $e$:
\begin{equation}
U_\text{B} =\kappa_\text{B}\sum_{e\in\mathcal{M}}(1-\mathbf{\hat{n}}_{t_a}\cdot \mathbf{\hat{n}}_{t_b})
\end{equation}
where $\kappa_\text{B}$ is the bending rigidity and $\mathbf{\hat{n}}_{t_a}$ and $\mathbf{\hat{n}}_{t_n}$ are the instantaneous normal vectors of $t_a$ and $t_b$, respectively. We relate tether strength and bending rigidity $\kappa_\text{T}=4\kappa_\text{B}$, observed for giant lipid vesicles \cite{VHA-VvBDAFGV2020}, to guarantee that the Young's modulus $Y=\frac{2}{\sqrt{3}}\kappa_T$ is always larger than the bending modulus $\kappa=\frac{\sqrt{3}}{2}\kappa_B$ \cite{SN1988,GK1996}.  Furthermore, we apply a local area conservation to each triangle by using a harmonic potential 
\begin{equation}
U_\text{A} =\frac{1}{2}\kappa_\text{A}\sum_{t\in\mathcal{M}}\frac{(A_{t_a}-A_0)^2}{A_0}
\end{equation}
with the area conservation coefficient $\kappa_\text{A}=50,000 k_BT$, the instantaneous triangle area $A_{t_a}$ and the target area for each triangle $A_0=\frac{A_\text{flex}}{N_t}=\frac{4\pi\cdot R^2}{3\cdot N_t}$. With this model, we assume that each flexicle conserves its surface area $A_\text{flex}$, yet is permeable to the surrounding solvent which allows the volume $V_\text{flex}$ to dynamically change.

%\begin{wrapfigure}{rt}{0.42\textwidth} 
%\end{wrapfigure} 

The active rod-shaped particles within the flexicles are modeled as rigid linear arrays of 5 overlapping spherical beads with diameter $\sigma$ (see \fig{fig:method}). We chose the aspect ratio $\alpha=3$ as such self-propelled rods reliably cluster at the membrane interface but are also significantly smaller than the size of the vesicle. The forces between bead centroids of different rods, between bead centroids and mesh vertices, and between non-neighboring mesh vertices are given by the purely repulsive Weeks-Chandler-Anderson potential:
\begin{equation}
U_\text{WCA}(r_{ij}) =
\begin{cases}
4\epsilon\left[\left(\frac{\sigma}{r_{ij}}\right)^{12}-\left(\frac{\sigma}{r_{ij}}\right)^{6}\right] - \epsilon &\text{if } r_{ij}<\sqrt[6]{2}\sigma\\
0 & \text{otherwise,}
\end{cases}
\end{equation}
with particle distance $r_{ij}$ and the energy unit $\epsilon=1$.

The equations of motion of each mesh vertex
\begin{equation}
\gamma_v \dot{\mathbf{r}}_i = \sum_j \mathbf{F}^\text{WCA}_{ij} + \mathbf{F}^\text{mesh}_{i} +  \sqrt{2\gamma_vk_BT}\mathbf{\eta}_i(t)
\end{equation}
and self-propelling rods
\begin{equation}
\begin{split}
\gamma_a \dot{\mathbf{r}}_i &= \sum_j \mathbf{F}^\text{WCA}_{ij} + F^\text{A} \hat{\mathbf{e}}_i +  \sqrt{2\gamma_ak_BT}\mathbf{\eta}_i(t)\\
\gamma_r \dot{\mathbf{\omega}}_i &= \sum_j \mathbf{T}^\text{WCA}_{ij} +  \sqrt{2\gamma_rk_BT}\mathbf{\xi}_i(t)
\end{split}
\end{equation}
follow Brownian dynamics where $\gamma_a=50$ and $\gamma_r=\frac{\sigma^3\gamma_a}{3}$ are the translational and rotational drag coefficients on the active rods, respectively. The coefficient $\gamma_v=5$ mimics the friction of a membrane embedded inside a viscous fluid according to the free-draining approximation \cite{GK1993}. While the excluded volume inter-particle forces $\mathbf{F}^\text{WCA}_{ij}$ and torques $\mathbf{T}^\text{WCA}_{ij}$ that act on both the mesh vertices and the active rods are given by the WCA potential, the inter-vertex forces $\mathbf{F}^\text{mesh}_{i}$ encompass all the mesh potentials mentioned above. The functions $\eta_i(t)$ and $\mathbf{\xi}_i(t)$ are normalized Gaussian white noise processes with zero mean $\langle \mathbf{\eta}_i(t)\rangle=0$, $\langle \mathbf{\xi}_i(t)\rangle=0$ , and unit variance $\langle \mathbf{\eta}_i(t)\mathbf{\eta}_j(t')\rangle=\delta_{ij}\delta(t-t')$, $\langle \mathbf{\xi}_i(t)\mathbf{\xi}_j(t')\rangle=\delta_{ij}\delta(t-t')$. The active force term $F^\text{A}_i \hat{\mathbf{e}}_i$ is only applied to the rod centers and points along the symmetry axis of the rod $\hat{\mathbf{e}}_i$ replicating Brownian motion. The active force magnitude $F^\text{A}=\frac{\text{Pe}\cdot k_BT}{\sigma}$ is controlled by the P\'eclet number $\text{Pe}$ of the active rods, which is a measure of translational activity.

For each ``state point'' we ran three independent simulations with $N_\text{flex}=3375$ flexicles ,which totals $N_\text{total}=3,459,375$ moving objects per simulation run, on 16 GPUs for a time period $t=200000\tau$ with time step $\Delta t = 0.001\tau$ and unit time $\tau=\sqrt{\frac{m\sigma^2}{k_BT}}$. Snapshots of each MD trajectory were stored at intervals of 250000 time steps. We chose to limit our parameter space to bending rigidities $\kappa_\text{B}\in[ 500 k_BT,\dots,25000 k_BT]$ and flexicle number densities $\rho_\text{flex}=\frac{4\pi}{3}{R^3}\frac{N_\text{flex}}{L^3}\in[0.3,\dots,1.1]$ to guarantee that all flexicles within the studied rigidity window are self-propelled with roughly the same effective velocity $v_\text{flex}$ and form an active gas at low densities. We used the open source molecular dynamics software HOOMD-blue [v3.1.0] to perform our simulations \cite{AGG2020}, the freud data analysis package for analysis \cite{RDHSAG2020} and the signac software package for data management \cite{ADRG2018}.

\section{Results}

\begin{figure*}[t!]
\centering
\includegraphics[width=\textwidth]{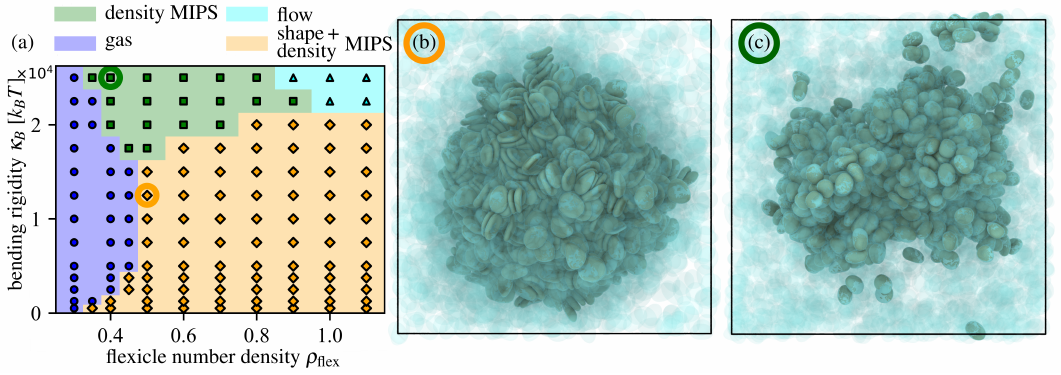}\\[0.2cm]
\includegraphics[width=\textwidth]{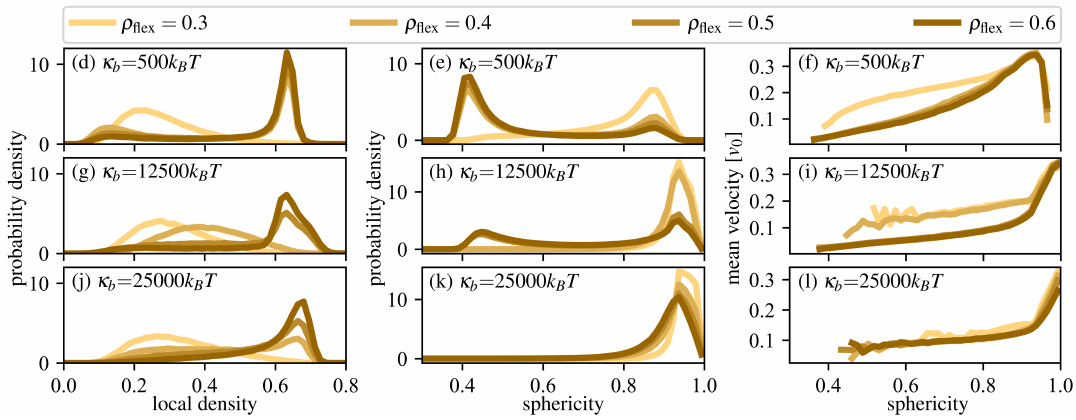}
\caption{Top: a) Steady-state phase diagram of $N_\text{flex}=3375$ flexicles each containing $N_\text{rod}=125$ rod-shaped, self-propelled particles with $\text{Pe}=200$. The parameter space is sampled based on the flexicle number density $\rho_\text{flex}=\frac{4\pi}{3}{R^3}\frac{N_\text{flex}}{L^3}$ and the bending rigidity $\kappa_\text{B}$. Simulation snapshots of a flexicle system in the b) combined shape and density MIPS and c) singular density MIPS steady-state regime. Only flexicles within the dense regions are depicted. Flexicles in the sparse, gas-like regions are shown as transparent. The orange and green circles indicate the position of both systems within the steady-state phase diagram. Bottom: Probability distributions of the local density (left column: d,g,j) and flexicle sphericity (center column: e,h,k) and mean flexicle velocity at different flexicle sphericities (right column, f,i,l) at different bending rigidities $\kappa_B$ (indicated in each panel) and flexicle number densities $\rho_\text{flex}$ (indicated by color) around the singular and combined MIPS transition.}
\label{fig:phase_diagram}
\end{figure*}

The collective behavior of flexicles is presented in a ``phase'' diagram of steady-state behavior (see \fig{fig:phase_diagram}) as a function of the membrane bending rigidity $\kappa_B$ and flexicle number density $\rho_\text{flex}$. Each vesicle contains $N_\text{rod}=125$ active rods with $\text{Pe}=200$. A ``phase'' diagram with $N_\text{rod}=83$ is depicted in Fig. SI1. Although the boundaries between the different steady-state behaviors change, qualitatively the same collective behaviors occur for both concentrations of rods $N_\text{rod}$. At low $\rho_\text{flex}$ the system forms a gas of self-propelled flexicles independent of $\kappa_B$. The self-propulsion mechanism of each flexicle is driven by the internal dynamics of the active rods, which spontaneously accumulate into a stable, polar dominant cluster at the membrane interface. The individual rods within the cluster align perpendicular to the membrane in accordance with their tendency to form smectic layers at hard walls \cite{WL2008,BGHP2020}. In this way, the rods collectively push against the vesicular wall, create a pocket of high Gaussian curvature, and propel the flexicle in the direction of the cluster's polar director as depicted in \fig{fig:method} (see also Movie 1). However, this propulsion mechanism breaks down by reducing either $N_\text{rod}<50$ or $\text{Pe}<50$ for the fixed flexicle size $R=8\sigma$ (see Fig. SI2). In these cases, either the internal rod density or the active force is insufficient to stabilize motility-induced clusters of rods. Additionally, it is known that the rods are prone to form multiple clusters that push in opposite directions when increasing $N_\text{rod}$ and significantly decrease the flexicle velocity \cite{UHB2023,LSG2023}. Consequently, the propulsion efficiency does not increase with $N_\text{rod}>125$.

Above a critical flexicle density, all systems undergo MIPS. However, we identified two different types of MIPS depending on the membrane bending rigidity: a combined shape and local density MIPS (see snapshot in \fig{fig:phase_diagram}b and Movie 2+3) and a simpler hard-particle-like local density MIPS (see snapshot in \fig{fig:phase_diagram}c and Movie 4). For low bending rigidities $\kappa_{B}<17000 k_BT$, we observe that the clustering of flexicles not only entails a separation into a dense jammed region of flexicles surrounded by a sparse fluid of flexicles (\fig{fig:phase_diagram}d+g) but also a coexistence of mostly spherical and highly aspherical flexicles as shown by their sphericity parameter $s=\frac{6\pi V_\text{flex}}{A_\text{flex}^{1.5}}$ in \fig{fig:phase_diagram}e+h) and the simulation snapshot in \fig{fig:phase_diagram}b. 
While the concentrated polar rod clusters cause a predominantly prolate shape for flexicles in the sparse region, flexicles in the dense region respond to the high collision pressure applied by neighboring flexicles and adopt oblate discotic shapes. These discs are arranged in small disordered stacks or rouleaux, alluding to the red blood-cell like shape of the flexicles and the similar structures observed in blood clots \cite{WSS2013}.

\begin{figure*}[t!]
\centering
\includegraphics[width=\textwidth]{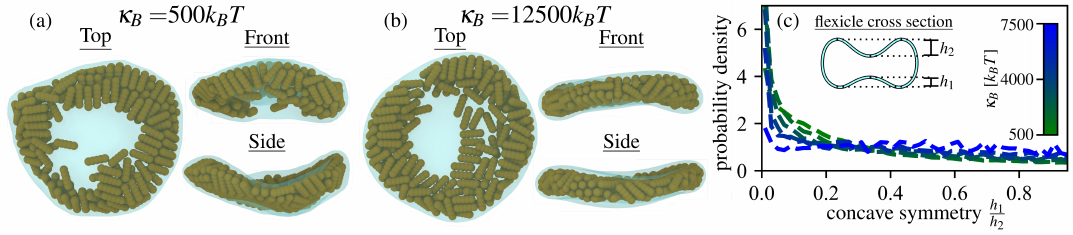}\\
\includegraphics[width=\textwidth]{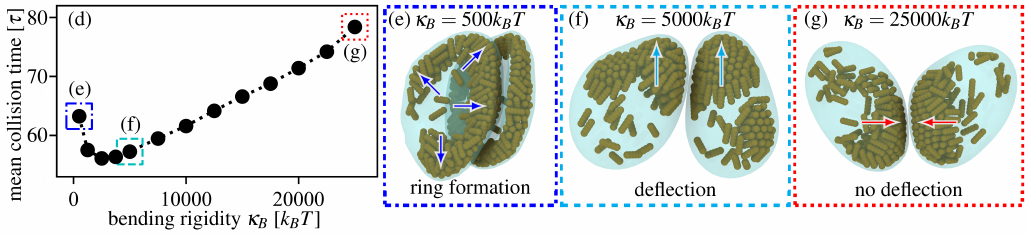}
\caption{Representative concave and biconcave discotic flexicle shapes observed in dense combined MIPS clusters for a) $\kappa_B=500k_BT$ and b) $\kappa_B=12500k_BT$, respectively. The internal rod-shaped particles form an apolar ring cluster with the particles pointing, on average, outward from the center of the flexicle. c) Probability distribution of the concave symmetry parameter of discotic flexicles during flexicle-flexicle collisions at flexicle number density $\rho_\text{flex}=0.3$. The concave symmetry parameter is defined as the ratio between the depths $h_2>h_1$ of the dips in the center of the flexicle on both sides of the disk as depicted in the sketch. d) Mean collision time between flexicles with different bending rigidities $\kappa_B$ in the gas phase at flexicle number density $\rho_\text{flex}=0.3$. Representative snapshots of two colliding flexicles that deform into disks and e) create rings of internal rods ($\kappa_b=500k_BT$), f) deflect their propulsion direction ($\kappa_b=5000k_BT$), and g) push against each other during the collision process without deflection ($\kappa_b=25000k_BT$). The arrows indicate the main propulsion direction of the flexicles.}
\label{fig:Transition}
\end{figure*}

By analyzing the dynamics of the self-propelled rods during the MIPS transition in the flexicle system see identify that its onset is coupled to the shape change in this regime. The mechanism for this combined MIPS can be understood as follows. As the discotic deformation of the membranes also modifies the internal particle confinement, the rods must occupy a space of significantly lower volume. The lower volume causes the compact propelling rod clusters, where each rod points in a similar direction, to become unstable. The rods change their internal arrangement and break the polarity within the deformed flexicle by forming rings along the interface (see \fig{fig:Transition}a+b). These rings are perpendicular to the shortest principal axis of the flexicle. The interruption of the propelling mechanism considerably slows down the flexicle velocity (see \ref{fig:phase_diagram}f+i) and enhances the collision time between flexicles, which is known to cause MIPS \cite{CT2015,BG2018}.  Furthermore, \fig{fig:Transition}c shows that in this regime the flexicle shapes within the dense cluster change from predominantly mono-concave bowl-like discs to a mixture with a considerable amount of highly biconcave, blood cell-like discocytes with increasing bending rigidity. This shape change indicates that upon compression, the flexicles adopt their energetically most favorable shape when the bending energies become dominant. Furthermore, we observe that the shape change into disks and the ring formation of the active rods can be triggered by flexicle-flexicle collisions for highly deformable flexicles (see \fig{fig:Transition}e), whereas similar deformation of more rigid flexicles requires multi-body crowding effects. Hence, two colliding flexicles with very low bending rigidity stay in contact for a longer time than more rigid flexicles, which only slightly deform and never lose the polarity of their internal rod clusters during collision (see \fig{fig:Transition}d). Consequently, the onset of the combined MIPS increases with $\kappa_{B}$ before it plateaus at $\rho_\text{flex}\approx0.49$ where the flexicles start to interact with multiple neighbors simultaneously.

Increasing $\kappa_B$ further, we observe that the flexicles undergo local density MIPS without significant shape changes (see snapshot in \fig{fig:phase_diagram}c). Here, the collision pressure between multiple particles is not high enough to sufficiently deform the membranes. Both in the dense and sparse regions the flexicles adopt a prolate, spheroidal shape with stable, polar internal clusters pushing the flexicle forward (see \fig{fig:phase_diagram}k+l). Consequently, the MIPS mechanism is similar to the one observed for hard self-propelled particles \cite{CT2015,BG2018}, where the collision process slows down the particles. Furthermore, we observe that in this regime flexicle rigidity increases the collision time (see \fig{fig:Transition}d), thereby promoting MIPS in accordance with earlier mean-field calculations of deformable active Brownian particles \cite{HLCMM2023}. In our system, however, we link the time of the collision process to a deflection mechanism where deformations can shift the pushing internal rod cluster, and consequently the propulsion direction of the flexicle, to the side as depicted in \fig{fig:Transition}f. The higher the deformation, the more the internal clusters redirect their polar direction from pointing toward the other flexicle to pointing toward an arbitrary side. This deflection of the flexicle propulsion direction facilitates the ability of flexicles to slide along each other more quickly and accelerates their separation. In contrast, a collision between rigid flexicles does not induce deformations that alter the propulsion direction of the flexicle. Hence, in these systems the flexicles push against each other throughout the collision process and stay pressed together for a longer time (see \fig{fig:Transition}g). The transition from the combined to the singular MIPS is also apparent in the global density of the system (see Fig.~SI3). When the system enters the combined MIPS state, the global density significantly decreases compared to the gas phase, as discotic flexicles within the dense cluster have a significantly lower volume than prolate flexicles in the gas. The density drop becomes weaker with increasing $\kappa_\text{B}$ until the global density becomes continuous and the flexicles enter the singular MIPS state where the particles in the dense and sparse region adopt similar prolate shapes.

\begin{figure*}[t!]
\centering
\includegraphics[width=\textwidth]{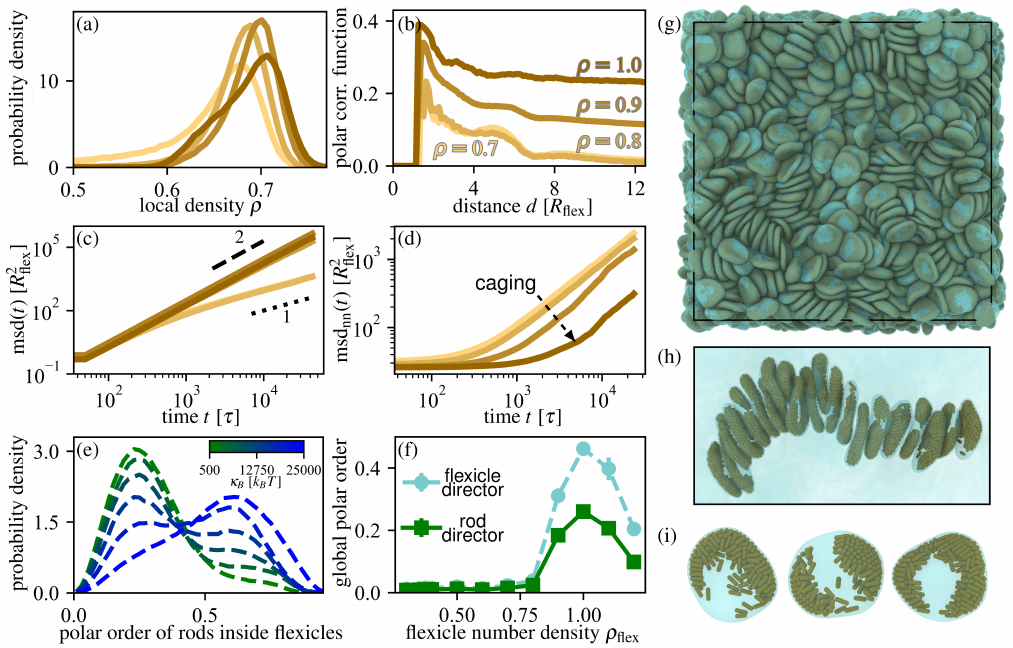}
\caption{a) Probability densities of the local density, b) polar correlation function, c) mean squared displacement $\text{msd}(t)= \langle(\mathbf{x}_\text{flex}(t)-\mathbf{x}_\text{flex}(0))^2\rangle$ of the center position of a flexicle $\mathbf{x}_\text{flex}(t)$ at time $t$, and d) mean squared distance $\text{msd}_\text{nn}(t)= \langle(\mathbf{r}_\text{ij}(t))^2\rangle$ of the relative flexicle distances $\mathbf{r}_\text{ij}$ between flexicle pairs $i$ and $j$ that are neighbors at time $t=0$. The bending rigidity for a)-d) is $\kappa_B=25000k_BT$. e) Probability distribution of the polar order of rods inside a flexicle for different $\kappa_B$ at flexicle number density $\rho_\text{flex}=1.0$. f) Global polar order of rod and flexicle propulsion directors at $\rho_\text{flex}=1.0$. g) Simulation snapshot of a flexicle system, h) representative rouleau stack and i) representative polar, discotic flexicle shapes in the jammed flow regime at $\rho_\text{flex}=1.0$ and $\kappa_B=25000k_BT$.}
\label{fig:flow}
\end{figure*}

At higher flexicle number density, rigid flexicles eventually must deform to avoid overlaps with their neighbors. Above the jamming transition, where the low-density tail of the local density distribution vanishes (see \fig{fig:flow}a), all flexicles deviate from a sphere shape (see \fig{fig:flow}g). and adopt morphologies reminiscent of biconcave red blood cell discocytes \cite{C1970,EM1991,H2016}. However, compared to flexicles with more deformable membranes in the combined MIPS state, rigid flexicles maintain a larger volume (see Fig.~SI3) such that the self-propelling rods within each flexicle have enough space to form non-closed rings along the membrane interface (see \fig{fig:flow}i). The rod clusters, therefore, do not lose polarity in rigid flexicles (see \fig{fig:flow}e) and push the flexicle in the direction of the cluster's polar director perpendicular to the shortest principal axis of the flexicles. Surprisingly, the combination of jamming and cluster polarity causes an emergent flow of neighboring flexicles apparent in the transition from diffusive to ballistic motion in \fig{fig:flow}c (see also Movie 5+6). An important aspect of this emergent flow is the observed rouleaux formation of the jammed flexicles (see \fig{fig:flow}h). 
As the principal axes of flexicles within the same rouleau align with the stacking direction, the polar rod clusters are perpendicular and, hence, are more likely to point in a similar direction. This local alignment of the internal rods and directors of self-propulsion between neighboring flexicles can be seen in the orientational correlation function even below the jamming transition (see \fig{fig:flow}b). In the jammed state, however, the rouleaux are long-lasting due to the associated reduced relative motion between neighboring flexicles as plotted in \fig{fig:flow}d. This can be seen as a caging effect that promotes the spontaneous alignment of the self-propulsion directions of multiple flexicles within the same rouleau. The local alignment eventually develops into a long ranged orientational ordering of the internal rods and the flexicle pushing directors $\mathbf{\hat{v}}_\text{flex}$ (see \fig{fig:flow}b+f), causing the whole system to flow collectively in an arbitrary direction. Additionally, we observe that the global alignment and the flow get weaker beyond an optimal flexicle number density. At these very high densities, the low flexicle volume constricts the motion of the internal self-propelled rods and prevents sufficient rod rearrangements.

\section{Discussion \& Conclusion}

In this paper, we studied the collective behavior of active, deformable vescles or "flexicles" across a range of membrane bending rigidities and flexicle number densities. We constructed a steady-state ``phase'' diagram that provides clear evidence for an intricate interplay between internal dynamics, shape transformations, and collective motion in these systems. First, we identified a bifurcation in the type of motility-induced phase separation states between highly flexible and more rigid flexicles. Flexicles with low bending rigidities not only cluster to form a dense region within a sparse fluid but also produce a coexistence of nearly spherical, free-moving flexicles and aspherical discotic shapes aggregating into small stacks. The accompaniment of deformations into aspherical shapes when flexicles enter a dense region is reminiscent of single cell organisms such as amoebae that respond to compression stresses inside dense collections by adopting elongated shapes \cite{RNSLL1999, CCTCN2013}. The clusters of discotic flexicles bear an even larger resemblance to red blood cell (RBC) clusters. Besides their nearly identical shape, RBCs are also known to create stacks or rouleaux in static or low-flowing blood plasma that facilitate the aggregation of blood clots \cite{WSS2013}. While the mechanisms of rouleaux formation in blood are not yet fully understood, two leading models have been proposed. One such model is the `bridging model', where macromolecules present in the surrounding fluid, such as fibrinogen or dextran, bond adjacent cells by adsorbing onto their cell membranes \cite{CS1987,B1988}. Another model explains the aggregation of RBCs as a result of depletion interactions \cite{KKL2021}. The presence of stacks in our systems of purely repulsively interacting flexicles underpins the idea that explicit attractive interactions are not required for rouleaux formation but rather reinforce their stability. Indeed, simulations of flexicles with short-range attractions ($<1.6\sigma$) between flexicle vertices ($N_\text{flex}=1000$, $\kappa_B\in[500k_BT,\dots,15000k_BT]$, and $\rho_\text{flex}=0.6$) confirm that the rouleaux are more stable with the addition of attraction independent of their analytical form (Gaussian, positive part of a Lennard-Jones, Yukawa potential, etc.). Lastly, we note that packings of flattened vesicular lipid membranes, also known as cisternae, play a vital role in the endoplasmic reticulum in the form of Golgi structures \cite{HW2017}. Although the assembly and disassembly of  Golgi membranes involves the fusion and separation of neighboring cisternae via oligomerization and dimerization of stacking proteins \cite{FYLLZHL2013}, their resemblance to flexicle stacks highlights that our flexicle model provides a promising basis to study the physics of even more complex biological processes.

Upon scrutinizing the internal dynamics of the rods within each flexicle during the combined MIPS transition, we related the collision-induced deformations of flexicles to the onset of motility-induced clustering. The morphological change of flexicles from spherical shapes to oblate disks impacts the confinement-dependent arrangement of rods, destabilizes the propulsive rod clusters, and fosters apolar ring-like formations, which decelerates the flexicles and dictates the onset of MIPS. The interdependence between internal rod distribution and membrane shape can be seen as a response mechanism to external stresses. The interplay between collision-induced external stresses, flexicle deformations, and modulated flexicle propulsions is analogous to feedback loops in cellular biological organisms where morphological shape change can influence internal dynamics of microtubules \cite{SSBRNT-SKGM2018,GSGDNC-MSKB2021} or actin \cite{SKCAA-GMDCLLJCPSS2019} and vice versa.

By increasing the membrane bending rigidity, we observed a transition from the combined shape and local density MIPS regime to a purely density-related phase-separated regime that is more similar to MIPS in hard-particle systems \cite{CT2015}. In the latter case, the collision pressures are insufficient to elicit the large deformations observed at higher membrane flexibilities, leading to MIPS that resembles the behavior of non-deformable, self-propelled spheres. However, beyond the jamming transition, these more rigid flexicles deform and adopt biconcave morphologies to prevent overlapping with neighbors. Here we discovered a spontaneous emergence of collective flow, which we ascribe to a combination of the internal rod clusters not forming closed rings and, hence, not losing their polarity, the formation of rouleaux structures that allow for the alignment of propulsion direction between neighboring flexicles, and the reduced relative motion between neighboring flexicles in the jammed state that promotes the propagation of forces within the system.

These three pillars of deformation, polar alignment, and jamming are also strategies that have been observed in cell migration and the collective behavior of cells within tissue \cite{SM2020,SUMTACC2010}. In particular, the relation between the motility and shape of cells has been studied extensively. Elongated cells, for instance, play a vital role in the fluidization and unjamming of cellular tissues frozen in place \cite{BYMM2016,MM2018,SI2024} and have been identified as markers for the dynamics within cancer cells \cite{GLOMSRGMFXPFWHBAMK2021} or within bronchial epithelial cells affected by asthma \cite{PKBMQTPMcGKGNSBRKTHSIWTHWMBDF2015}. Another key feature of cell migration is the polar alignment of neighboring cells \cite{EJ2011,RPCG-MABS2011}. One process that is particularly similar to our observations in collective flexicle systems is known as contact inhibition of locomotion. Here, the cells redirect and align their motion due to the deformations inflicted by their collision \cite{MO2012,CCTCN2013,SM2017}. Lastly, the collective flexicle flow is nearly identical to active jamming \cite{HFM2011} or `solid flocking' phenomena \cite{MCGBLLDFOMLDBPOUTPMYFCS2017, GPMBSMCM2018} in epithelial cells. Like in our flexicle system, the slowed dynamics and jamming between neighboring aligning cells results in the collective motion of an otherwise solidified cell front.

In conclusion, our study of flexicles opens a pathway to design new classes of responsive microswimmers and materials with autonomous emergent behaviors and motions. Additionally, while superficial, the striking dynamical similarity between flexicles and biological cells makes our flexicle model also a promising candidate for designing artificial tissues or studying even more complex biological processes. Although hydrodynamic effects that may be present in flexicles were neglected in this study, we expect that additional interactions will only alter our phenomenological observations quantitatively as long as the internal self-propelled particles can form a polar cluster, pushing the flexicles forward.

\section*{Acknowledgements}
The authors thank Sophie Y. Lee for helpful discussions. We were supported by a grant from the Simons Foundation (256297, SCG). An award of computer time was provided by the INCITE program. This research used resources from the Oak Ridge Leadership Computing Facility, which is a DOE Office of Science User Facility supported under Contract DE-AC05-00OR22725. This research was also supported in part through computational resources and services provided by Advanced Research Computing at the University of Michigan, Ann Arbor.

\bibliography{reference} % Entries are in the refs.bib file

\end{document}

\title{Supplementary information:\\Collective behavior of ``flexicles''}%

\author{Philipp W. A. Sch\"onh\"ofer$^1$ and Sharon C. Glotzer$^{1,2}$}%
\email{Philipp W. A. Sch\"onh\"ofer (pschoenh@umich.edu), Sharon C. Glotzer (sglotzer@umich.edu)}
\affiliation{Department of Chemical Engineering, University of Michigan, Ann Arbor, Michigan 48109, USA.}
\affiliation{Biointerfaces Institute, University of Michigan, Ann Arbor, Michigan 48109, USA.}

\maketitle

\section*{Electronic Supplementary Information}

\begin{itemize}
\item Movie 1: \verb"single_flexicle.mov" Video of a single flexicle with bending rigidity $\kappa_B=25000k_BT$ and $N_\text{rod}=125$ internal self-propelled rods with P\'eclet number Pe$=200$. The colloidal rods spontaneously cluster at the vesicle interface, align against the wall, and push the flexicle forward.
\item Movie 2: \verb"combined_mips_500.mov" Video of the combined shape and local density motility-induced phase separated steady state of $N_\text{flex}=3375$ flexicles with bending rigidity $\kappa_B=500k_BT$ and $N_\text{rod}=125$ internal self-propelled rods with P\'eclet number Pe$=200$ at flexicle number density $\rho_\text{flex}=0.4$.
\item Movie 3: \verb"combined_mips_12500.mov" Video of the combined shape and local density motility-induced phase separated steady state of $N_\text{flex}=3375$ flexicles with bending rigidity $\kappa_B=12500k_BT$ and $N_\text{rod}=125$ internal self-propelled rods with P\'eclet number Pe$=200$ at flexicle number density $\rho_\text{flex}=0.5$.
\item Movie 4: \verb"singular_mips_25000.mov" Video of the singular local density motility-induced phase separated steady state of $N_\text{flex}=3375$ flexicles with bending rigidity $\kappa_B=25000k_BT$ and $N_\text{rod}=125$ internal self-propelled rods with P\'eclet number Pe$=200$ at flexicle number density $\rho_\text{flex}=0.4$. Flexicles in the sparse region are transparent.
\item Movie 5: \verb"flow_25000.mov" Video of the emergent flow steady state of $N_\text{flex}=3375$ flexicles with bending rigidity $\kappa_B=25000k_BT$ and $N_\text{rod}=125$ internal self-propelled rods with P\'eclet number Pe$=200$ at flexicle number density $\rho_\text{flex}=1.0$.
\item Movie 6: \verb"flow_removed_drift_25000.mov" Video of the emergent flow steady state of Movie 5 with the center of mass drift removed.
\end{itemize}

\section*{Steady-state phase diagram of flexicles with smaller loading of internal self-propelled rods}

\begin{figure}[h!]
\centering
\includegraphics[width=0.75\textwidth]{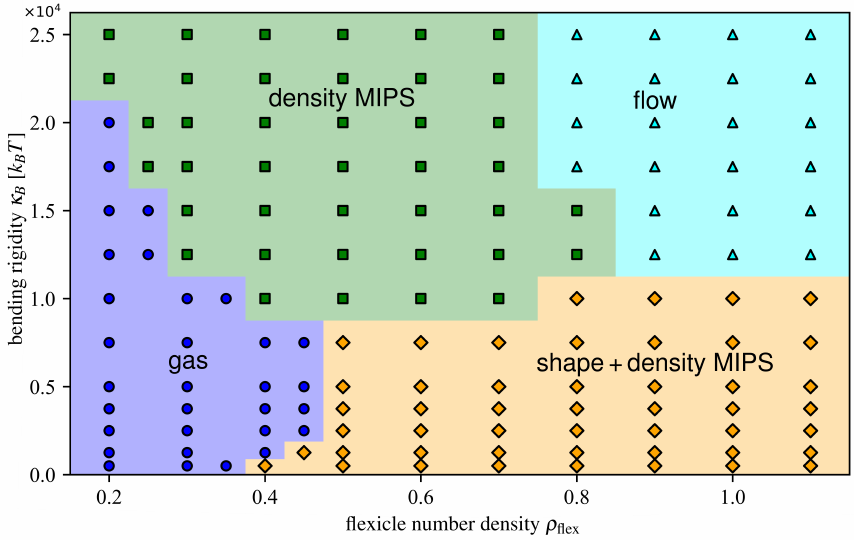}
\caption{Steady-state phase diagram of $N_\text{flex}=3375$ flexicles each containing $N_\text{rod}=83$ rod-shaped, self-propelled particles with $\text{Pe}=200$. The parameter space is sampled based on the flexicle number density $\rho_\text{flex}=\frac{4\pi}{3}{R_\text{flex}^3}\frac{N_\text{flex}}{L^3}$ and the bending rigidity $\kappa_\text{B}$.}
\label{fig:phase_diagramSI}
\end{figure}

\section*{Mean velocity of single flexicles at different bending rigidities and loading of internal self-propelled rods}

\begin{figure}[h!]
\centering
\includegraphics[width=\textwidth]{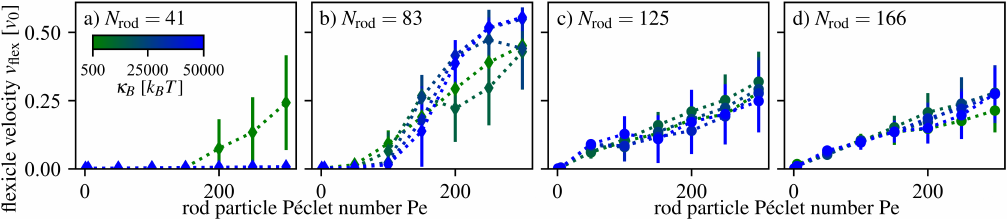}
\caption{Mean velocity $v_\text{flex}$ of single flexicles for different bending rigidities $\kappa_B$ and P\'eclet number of the internal self-propelled rods. The number of rods $N_\text{rod}$ inside the flexicle is indicated in the pannels. The velocities are averaged over 100 replicates.}
\label{fig:single_flexicle}
\end{figure}

\section*{Flexicle volumes at different bending rigidities and flexicle number densities}

\begin{figure}[h!]
\centering
\includegraphics{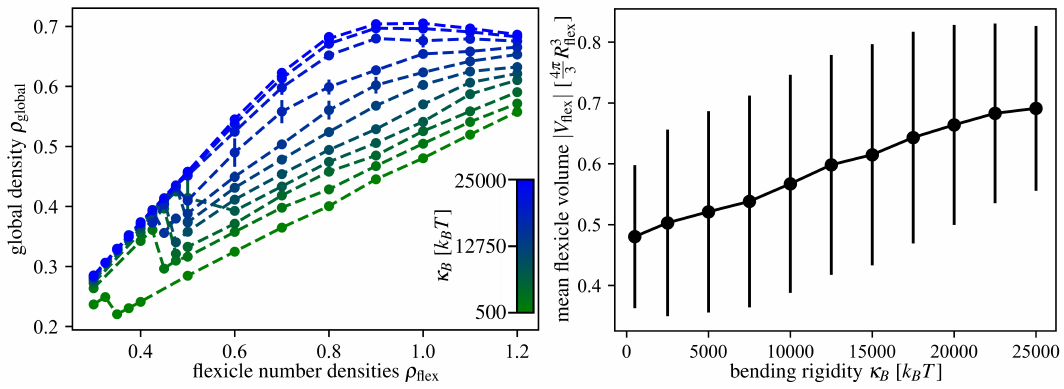}
\caption{Left: Global density $\rho_\text{global}=\frac{\sum^{N_\text{flex}}_i V_i}{L^3}$ of the flexicle system at different flexicle number densities $\rho_\text{flex}=\frac{4\pi}{3}{R_\text{flex}^3}\frac{N_\text{flex}}{L^3}$ and bending rigidities $\kappa_B$. Here, $N_\text{flex}=3375$ is the number of flexicles in the simulation box, $V_i$ is the instantaneous volume of flexicle $i$, $L$ is the simulation box length and $R_\text{flex}$ is the radius of a non-deformed spherical flexicle. The drop in $\rho_\text{global}$ indicates the transition from the active gas into the combined MIPS state that includes flexicle deformations into disks. The transition from active gas into the mono motility-induced phase-separated state does not feature significant morphological changes to the flexicle and, hence, does not create a drop in $V_i$. Right: Mean flexicle volume in multi-flexicle systems at $\rho_\text{flex}=1.0$ at different bending rigidities $\kappa_B$.}
\label{fig:global_density}
\end{figure}